\documentclass[%
 aip,
 amsmath,amssymb,
 reprint,%
]{revtex4-1}
\usepackage[utf8]{inputenc}
\usepackage{graphicx}
\begin{document}

\title{Many-particle Quantum Hydrodynamics of  Spin-1 Bose-Einstein Condensates}

\author{Mariya Iv. Trukhanova}
\email{trukhanova@physics.msu.ru}
 \affiliation{M. V. Lomonosov Moscow State University, Faculty of Physics, Leninskie Gory,  Moscow, Russia}
\affiliation{Russian Academy of Sciences, Nuclear Safety Institute,
B. Tulskaya 52, 115191 Moscow, Russia}
\author{Yuri N. Obukhov}
\affiliation{Russian Academy of Sciences, Nuclear Safety Institute,
B. Tulskaya 52, 115191 Moscow, Russia}

\begin{abstract}
We develop a novel model of the magnetized spin-1 Bose-Einstein condensate (BEC) of neutral atoms,  using the method of many-particle quantum hydrodynamic (QHD) and propose an original derivation of the system of continual equations. We consider bosons with a spin-spin interaction and a short range interaction in the first order in the interaction radius, on the of basis of the self-consistent field approximation of the QHD equations. We demonstrate that the dynamics of the fluid velocity and magnetization is determined by a nontrivial modification of the Euler and Landau-Lifshitz equation, and show that a nontrivial modification of the spin density evolution equation contains the spin torque effect that arises from the self-interactions between spins of the bosons. The properties of the dispersion spectrum of collective excitations are described. We obtain the new contribution of the self-interaction of spins in the spin wave spectrum together with the influence of an external magnetic field and spin-spin interactions between polarized particles. The  anisotropic spin wave instability is predicted.
\end{abstract}

%
%
%
\maketitle
\tableofcontents
%
%

\section{Introduction}

Magnetized Bose-Einstein condensates (BEC) are in the focus of numerous studies. The dynamics of the scalar Bose-Einstein condensate had been analyzed in details in the mean-field approach by Gross \cite{1} and Pitaevskii \cite{2}. Recently, the attention has been growing to the study of BEC for atoms with magnetic polarization, due to the experimental realization\cite{100} of the polarized magnetic BEC of $^{52}$Cr, and to the study of various non-trivial magnetic structures in the BEC, that can exist in a spinor Bose-Einstein condensate.

In a system of cold atoms with spins in the BEC state, a number of interesting exotic topological excitations such as vortices and skyrmions \cite{3,4,5} can exist. Skyrmions are investigated in many condensed matter systems, such as liquid crystals \cite{50}, bulky magnetic materials and thin films \cite{51}, as well as the quantum Hall systems \cite{52}. The stability of skyrmions in a fictitious spin-${\frac 12}$ condensate of $^{87}$Rb atoms had been studied in Ref. \cite{6}. It was found that skyrmions can exist in the two-component spinor only as a metastable state, and in addition, the size and the lifetime of the skyrmion, its spin texture and its dynamical properties were obtained.  The two-dimensional skyrmions were investigated theoretically \cite{7} in a spin-$1$ BEC and realized for the first time experimentally \cite{8} in a spin-$2$ Bose-Einstein condensate. The nonequilibrium dynamics of a rapidly quenched spin-$1$ Bose gas with spin-orbit coupling had been studied \cite{9} by solving the stochastic projected Gross-Pitaevskii equation. It was shown that the crystallization of half-skyrmions (merons) can be occur in a spinor condensate of $^{87}$Rb atoms. The three-dimensional skyrmions, in the general case, are unstable towards shrinkage due to an energy gradient. But the new way of creation of the three-dimensional skyrmions in a ferromagnetic spin-$1$ BEC by manipulating a multipole magnetic field and a pair of counter-propagating laser beams had been proposed recently \cite{10}.

The hydrodynamical description of the BEC is based on the continuum mechanics concepts such as the density of the particles, the mass current density as well as the force fields densities. The construction of the hydrodynamical description of a spinor condensate helps to simplify the consideration of the dynamics of different exotic topological excitations such as skyrmions, merons and vortices. A set of hydrodynamical equations had been obtained recently \cite{11} for a spin-$1$ BEC, which are equivalent to the multi-component Gross-Pitaevskii equations, but in terms of the observable physical quantities such as the spin density and the nematic (or quadrupolar) tensor in addition to the density and the mass current. The hydrodynamical equations of motion for a ferromagnetic BEC of arbitrary spin in the long wavelength limit were derived by A. Lamacraft \cite{110}. A nontrivial modification of the Landau-Lifshitz equation  was obtained, in which the magnetization is defined by the superfluid velocity, and the anisotropic spin wave instability was discussed. The mean-field hydrodynamical equations for the dynamics of a spin-F spinor condensates had been also derived in Ref. \cite{12}, where the generalization to spin-$1$ of the Mermin-Ho relation was proposed and an analytic solution for the skyrmion texture in the incompressible regime of a spin-half condensate had been found. The three-fluid hydrodynamical description had been developed in Ref. \cite{13} to treat the low frequency and long wavelength excitations of the spin-$1$ BEC. The hydrodynamical formulation of the mean-field dynamics of a system, derived in Refs. \cite{11,110,12,13}, is based on the introduction of a second-quantized Hamiltonian.  The well-known mean-field method cannot take into account inter-particle interactions, based on the collective dynamics of individual particles, and it also does not take into account thermal fluctuations in a system of a large number of particles. Although the mean-field method makes it possible to obtain a hydrodynamical model of a Bose condensate of ultracold atoms, it cannot take into account thermal fluctuations and inter-particle interactions in an explicit form.

In this article we revisit the original derivation of the equations of quantum hydrodynamics of BEC on the basis of the collective dynamics of individual quantum particles described by the many-particle Schr\"{o}dinger equation. Our aim is to derive the macroscopic equations for the evolution of physical fields, and to confirm and expand the previous results by making use of the many-particle quantum hydrodynamics method.

The method of many-particle quantum hydrodynamics makes it possible to pass from a description in terms of a many-particle wave function of the Schr\"{o}dinger equation to the hydrodynamical equations formulated in the terms of macroscopic variables such as the particles density, the mass current density, the spin density and the spin current density. This allows one to study multiparticle interactions and nonequilibrium processes in a system of a large number of interacting particles. The method of many-particle quantum hydrodynamics was developed for various physical systems such as the quantum plasmas \cite{14,15}, the spinning plasmas \cite{16,17}, the relativistic quantum plasma \cite{18}, ultracold Bose and Fermi gases with the short range interaction \cite{19,20} and magnetized BEC \cite{21}. Andreev \cite{21} established the equations of the quantum hydrodynamics (QHD) for the magnetized spin-$1$ neutral BEC, where the spin-spin interaction along with the short range interaction had been taken into account.

This paper is organized as follows. In Sec. \ref{I} we present the derivation of the set of QHD equations for a ferromagnetic Bose condensate of spin-$1$ with the help of the method of many-particle quantum hydrodynamics. In Sec. \ref{III} we demonstrate the derivation of the macroscopic equations of the hydrodynamics of a Bose condensate, based on the explicit representation of the many-particle wave function. In Sec. \ref{II} we analyze the small excitations in the magnetized BEC.

\section{ \label{I} General theory}

\subsection{\label{I1}Euler angles description}

When the magnetization vector of the ferromagnetic BEC is directed along the $z$-axis, the order parameter is described by $\psi_{(0)} = \left(\begin{array}{c}1\\ 0\\ 0\end{array}\right)$. The mean-field state of the BEC can be determined by the five variables: the density $\rho$, the phase $\xi$  and, the $\hat{U}{}^1$ gauge transformation that depends on the three Euler angles $\varphi,\vartheta,\chi$. A general order parameter $\psi$ is given by performing a gauge transformation with the phase $\xi$ and rotating to an arbitrary direction specified by the Euler angles  $\varphi,\vartheta,\chi$ \cite{11}
\begin{equation}
\psi=\begin{pmatrix} \psi_{1} \\ \psi_{0} \\ \psi_{-1}\end{pmatrix}=\sqrt{\rho}e^{i\xi}\hat{U}^1(\varphi,\vartheta,\chi)\begin{pmatrix} 1 \\ 0 \\ 0\end{pmatrix}
\end{equation}
When rotating the coordinate system by the Euler angles $\varphi,\vartheta,\chi$, the spin functions and the spin matrices for the spin $F=1$ are transformed via the rotation operator
\begin{equation}
\hat{U}^1(\varphi,\vartheta,\chi)=e^{-i\varphi\hat{F}_z}e^{-i\vartheta\hat{F}_y} e^{-i\chi\hat{F}_z},
\end{equation}
and in the representation of the cyclic basis, the operator $\hat{U}^1$ has the form
\begin{equation}
\hat{U}^1=\begin{pmatrix} \cos^2\frac{\vartheta}{2}e^{-i(\varphi+\chi)} & -\frac{\sin\vartheta}{\sqrt{2}}e^{-i\varphi} & \sin^2\frac{\vartheta}{2}e^{-i(\varphi-\chi)} \\\frac{\sin\vartheta}{\sqrt{2}}e^{-i\chi} & \cos\vartheta & -\frac{\sin\vartheta}{\sqrt{2}}e^{i\chi} \\ \sin^2\frac{\vartheta}{2}e^{i(\varphi-\chi)}& \frac{\sin\vartheta}{\sqrt{2}}e^{i\varphi} &\cos^2\frac{\vartheta}{2}e^{i(\varphi+\chi)} \end{pmatrix}.
\end{equation}
In this way, the general order parameter of the ferromagnetic BEC is given by \cite{11}
\begin{equation} \label{wave}
\begin{pmatrix} \psi_{1} \\ \psi_{0} \\ \psi_{-1}\end{pmatrix}=\sqrt{\rho}e^{i(\xi-\chi)}\begin{pmatrix} \cos^2\frac{\vartheta}{2}e^{-i\varphi} \\ \frac{\sin\vartheta}{\sqrt{2}} \\ \sin^2\frac{\vartheta}{2}e^{i\varphi} \end{pmatrix}=\sqrt{\rho}e^{i\xi}\mathbf{z}. \end{equation}
The order parameter (\ref{wave}) is normalized to the number of particles $N$
\begin{equation}
\int d^3\mathbf{x}\,\psi^{\dagger}\psi = N,
\end{equation}
and the spinor $\mathbf{z}$ is normalized to unity $\mathbf{z}^{\dagger}\mathbf{z}=1$.

\subsection{Many-particle quantum hydrodynamics description}

Let us consider a system of $N$ bosonic neutral atoms with spin-1 and a short-range potential. The microscopic quantum dynamics of the system of bosons is described by the many-particle Schr\"{o}dinger equation with the Hamiltonian
\begin{eqnarray}
\hat{H}=\sum_{j=1}^{N}\biggl(\frac{1}{2m_j}\hat{p}^\alpha_j\hat{p}_{j\alpha}+U_{j,ext}(\mathbf{r}_j,t)
-\gamma_j\hat{f}^{\alpha}_jB_{\alpha}\biggr)\nonumber\\ \label{H}
+\,\frac{1}{2}\sum_{j,k,j\neq k}^{N}\biggl(U_{int}(|\mathbf{r}_{jk}|)-\gamma_j\gamma_k
G^{\alpha\beta}_{jk}(|\mathbf{r}_{jk}|)\hat{f}^{\alpha}_j\hat{f}^{\beta}_k\biggr),
\end{eqnarray}
where $\hat{p}^\alpha_j$ is the momentum operator of the $j$-th atom, $m_j$ is the mass of the particle, $\gamma_j=\gamma$ is the gyromagnetic ratio, takes the same value for all particles of the system, $U_{j,ext}$ is the potential energy in the external force field or a spin-independent potential such as an optical confinement trap, $U_{int}$  is the short-range interaction potential which goes to zero at large inter-particle distances $|\mathbf{r}_{jk}|$. The first term in the Hamiltonian represents the kinetic energy operator and the third term is the linear Zeeman energy of $j$-th particle. The last term represents the spin-spin interactions between polarized bosons. The spin-spin interactions between atoms are represented by the Green function
\begin{equation}
G^{\alpha\beta}_{jk}(|\mathbf{r}_{jk}|)=4\pi\delta^{\alpha\beta}\delta_{jk}+\partial^{\alpha}_j
\partial^{\beta}_k(\frac{1}{|\mathbf{r}_{jk}|}).
\end{equation}
The Hamiltonian of the ferromagnetic Bose-Einstein condensate (\ref{H}) determines the quantum dynamics of the system from the Schr\"{o}dinger equation
\begin{equation} \label{Shrodinger}
  i\hbar\frac{\partial\psi_a(R,t)}{\partial t}=\hat{H}\psi_a(R,t),
\end{equation}
where the many-particle wave function of the condensate
\begin{equation}
\psi_a(R,t)=\psi_a(\mathbf{r}_1, \mathbf{r}_2,...,\mathbf{r}_N,t).
\end{equation}
is a spinor function in $3N$ configuration space, and $a$ is the spin index $(a=-F,...,F)$.

The state of the condensate is characterized by the density in the neighborhood of $\mathbf{r}$ in a physical space as
\begin{equation} \label{n}
  \rho(\mathbf{r},t)=\int dR \sum_{j=1}^N\delta(\mathbf{r}-\mathbf{r}_j)\psi^*_a(R,t)\psi_a(R,t)=\langle\psi^{\dagger}\psi\rangle.
\end{equation}
The concentration function $\rho(\mathbf{r},t)$ is thus determined as the quantum average of the concentration operator $\hat{\rho}=\sum_j\delta(\mathbf{r}-\mathbf{r}_j)$ in the coordinate representation.
The spin density vector of the polarized bosons is determined in a similar way
\begin{equation}  \label{F3}
  F_{a}(\mathbf{r},t)=\int dR \sum_{j=1}^N\delta(\mathbf{r}-\mathbf{r}_j)\psi^*_b(R,t)(\hat{f}_{j,a})_{bc}\psi_c(R,t),\end{equation}
as the quantum average of the spin operator, where the spin matrices $\hat{\mathbf{f}}_j$ of spin-1 particles can be written as
\begin{equation} \label{spinmatrix}
\left. \begin{aligned}\hat{f}_x &=\frac{1}{\sqrt{2}}\begin{pmatrix} 0 & 1 & 0 \\ 1 & 0 & 1 \\ 0 & 1 & 0 \end{pmatrix},\\
\hat{f}_y &=\frac{1}{\sqrt{2}}\begin{pmatrix} 0 & -i & 0 \\ i & 0 & -i \\ 0 & i & 0 \end{pmatrix},\\
\hat{f}_z &=\begin{pmatrix} 1 & 0 & 0 \\ 0 & 0 & 0 \\ 0 & 0 & -1 \end{pmatrix}.\end{aligned}\qquad\right\}
\end{equation}

\subsection{\label{I2}The macroscopic quantum hydrodynamics equations}

The continuity equation for the concentration of particles $\rho(\mathbf{r},t)$ can be derived by taking the time derivative of the definition of concentration (\ref{n}) and applying the many-particle Schr\"{o}dinger equation
\begin{equation}\label{nn}
  \partial_t \rho(\mathbf{r},t)+\partial_{\alpha}J^{\alpha}(\mathbf{r},t)=0,
\end{equation}
where the current density has the microscopic definition
\begin{equation} \label{current}
 \mathbf{J}(\mathbf{r},t)=\biggl\langle\frac{1}{2m_j}\biggl(\psi^{\dagger}\hat{\mathbf{p}}_j\psi+
 (\hat{\mathbf{p}}_j\psi)^{\dagger}\psi\biggr)(R,t)\biggr\rangle.
\end{equation}
The evolution equation for the current (\ref{current}) is obtained by taking the time derivative of the definition for the current density (\ref{current}) and invoking the Schr\"{o}dinger equation (\ref{Shrodinger})
\begin{widetext}
\begin{equation}\label{Euler}
m\frac{\partial J^{\alpha}}{\partial t}+\partial_{\beta}\Pi^{\alpha\beta}=\gamma F_{\beta}\partial^{\alpha}B^{\beta}-
\rho\partial^{\alpha}U_{ext}-\int d\mathbf{r}'\partial^{\alpha}U_{int}(\mathbf{r},\mathbf{r}')\rho_2(\mathbf{r},\mathbf{r}',t)+\int d\mathbf{r}'\partial^{\alpha}G_{\mu\nu}(\mathbf{r},\mathbf{r}')F^{\mu\nu}_2(\mathbf{r},\mathbf{r}',t).
\end{equation}
\end{widetext}   In the equation of motion (\ref{Euler}) $F^{\beta}$ is the spin density and $B^{\beta}$ is the total magnetic field.
The tensor $\Pi^{\alpha\beta}$  on the left hand side of equation (\ref{Euler}) is the momentum flux tensor   \begin{equation}\label{Flux} \Pi^{\alpha\beta}(\mathbf{r},t)=\frac{1}{4m}\biggl\langle \psi^{\dagger}\hat{p}^{\alpha}_j\hat{p}^{\beta}_j\psi+(\hat{p}^{\alpha}_j\psi)^{\dagger}\hat{p}^{\beta}_j\psi+h.c.\biggr\rangle. \end{equation}
The terms on the right hand side of equation (\ref{Euler}) are the force fields due to inter-particle interactions and the action of external fields. The inter-particle short-range interaction is comparable with the radius of bosons. The main contribution in the force field of the short-range interactions comes in the first order in the small parameter (the interaction radius). Interactions are determined by the two-particle correlation functions, which can be represented in the form of
\begin{equation}   \label{rho2}
\rho_2(\mathbf{r},\mathbf{r}',t)=\int dR\sum_{j,k,j\neq k} \delta(\mathbf{r}-\mathbf{r}_j)\delta(\mathbf{r}'-\mathbf{r}_k)\psi^{\dagger}(R,t)\psi(R,t).
\end{equation}
The spin-spin interactions are described by the last term of the equation of motion (\ref{Euler}), where the two-spins correlations are characterized by the function
\begin{equation}  \label{F2}
F^{\mu\nu}_2(\mathbf{r},\mathbf{r}',t)=\int dR\sum_{j,k,j\neq k} \delta(\mathbf{r}-\mathbf{r}_j)\delta(\mathbf{r}'-\mathbf{r}_k) \gamma_j\gamma_k\psi^{\dagger}\hat{f}^{\mu}_j\hat{f}^{\nu}_k\psi,
\end{equation}
where $\hat{f}^{\mu}_j$ are given by (\ref{spinmatrix}).

\section{ \label{III} The Madelung decomposition}

The Madelung decomposition of the $N$-particle wave function can be formulated in terms of the amplitude $a(R,t)$, the phase $\xi(R,t)$  and the local spinor, defined in the local frame of reference, normalized as $\mathbf{z}^{\dagger}\mathbf{z}=1$:
\begin{equation} \label{wave3}
  \psi(R,t)=a(R,t)\exp \biggl(\frac{i}{\hbar}\xi(R,t)\biggr)\mathbf{z}(R,\mathbf{r},t).
\end{equation}
Applying the decomposition (\ref{wave3}) for the $j$-th particle, a microscopic superfluid velocity can be introduced by
\begin{equation}
  v_j^\alpha(R,\mathbf{r},t)=\frac{\hbar}{m_j}(\nabla_j^\alpha\xi-i\mathbf{z}^{\dagger}\nabla_j^\alpha\mathbf{z}),
\end{equation}
and in the terms of Euler angles
\begin{equation}
v_j^\alpha=\frac{\hbar}{m_j}\Biggl(\nabla_j^\alpha(\xi-\chi) - \cos\vartheta\nabla_j^\alpha\varphi\Biggl).
\end{equation}
After the substitution of the wave function in the explicit form (\ref{wave3}), the microscopic momentum flux tensor (\ref{Flux}) is recast into
\begin{widetext}
  \begin{equation}\label{Flux2}
\Pi^{\alpha\beta}(\mathbf{r},t)= \biggl\langle\frac{\hbar^2}{2m}(\partial_j^{\alpha}a\partial_j^{\beta}a-a\partial_j^{\alpha}\partial_j^{\beta}a)
+a^2m v^{\alpha}_jv^{\beta}_j+\frac{\hbar^2}{2m}a^2\partial_j^{\alpha}\mathbf{f}_j\cdot\partial_j^{\beta}\mathbf{f}_j\biggr\rangle,
  \end{equation}
\end{widetext}
where the first term in the microscopic definition of the momentum flux tensor is the Bohm potential, the second term is the  fluid pressure and the last term is the ``spin stress'', produced by the self-interaction of the spin particles, or in other words, the spin part of the Bohm potential.

We split the velocity field of the $j$-th particle as $v^{\alpha}_j(R,\mathbf{r},t)=z^{\alpha}_j(R,\mathbf{r},t)+v^{\alpha}(\mathbf{r},t)$, where $z^{\alpha}_j(R,\mathbf{r},t)$ is the thermal part of the velocity or the fluctuations velocity about the macroscopic average $v^{\alpha}$. In a similar way, the spin of the $j$-th particle can be represented as $f^{\alpha}_j(R,\mathbf{r},t)=w^{\alpha}_j(R,\mathbf{r},t)+f^{\alpha}(\mathbf{r},t)$, where $w^{\alpha}_j$ describes the spin fluctuations about the macroscopic average $f^{\alpha}$. The particle system is assumed to be closed and not placed in a thermostat. The temperature is obtained from the average kinetic energy of the chaotic motion of atoms of our system. In accordance with this definition, we introduce deviations of the velocity and the spin of quantum particles from the local average values $z^{\alpha}_j$ and $w^{\alpha}_j$, which correspond to the ordered motion of the particles.  Substituting these expressions into the definition of the momentum flux density (\ref{Flux2}), we can obtain the contribution of the kinetic pressure and the spin-thermal effects into the dynamics of the particles.

Following the conclusion of the Ref. \cite{22}, we write the inter-particle interaction of bosons for the finite temperature, represented by the third term of the equation (\ref{Euler}),  in the form
\begin{eqnarray}
-\,\frac{1}{2}\int dR\sum_{j,k,j\neq k}\left\{  \delta(\mathbf{r}-\mathbf{r}_j)-
\delta(\mathbf{r}'-\mathbf{r}_k) \right\}\times\nonumber\\
\times\,\partial^{\alpha}U_{int}(\mathbf{r}_{jk})\psi^{\dagger}\psi,\qquad
\mathbf{r}_{jk}=\mathbf{r}_{j}-\mathbf{r}_{k}.
\end{eqnarray}
On the basis of the results obtained in \cite{21} we can represent the density of the interaction force for bosons with a short-range interaction potential in the form of the divergence of the tensor field $\sigma^{\alpha\beta}(\mathbf{r},t)$, which is the quantum stress tensor determined by the inter-particle interaction \cite{21}. The  inter-particle interaction potential for the system of bosons close to the BEC state can be finally derived in the form
\begin{equation}
\sigma^{\alpha\beta}(\mathbf{r},t)=-\frac{1}{2}\Upsilon\delta^{\alpha\beta} \biggl(2\rho_{bec}\rho_n+2\rho^2_n+\rho^2_{bec}   \biggr),
\end{equation}
where $\rho_{bec}$ denotes the concentration of particles in the BEC state and $\rho_n$ is the concentration of excited particles. The constant $\Upsilon$ represents the short range interaction at first order in the interaction radius and is determined by the following integral
\begin{equation}
  \Upsilon=\frac{4\pi}{3}\int dr\,r^3\,\frac{\partial U_{int}(r)}{ \partial r}.
\end{equation}
In the self-consistent  field approximation two-particle correlation functions (\ref{rho2}) and (\ref{F2}) can be represented as
\begin{eqnarray}
\rho_2(\mathbf{r},\mathbf{r}',t) &=& \rho(\mathbf{r},t)\rho(\mathbf{r}',t),\\
F_2^{\alpha\beta}(\mathbf{r},\mathbf{r}',t) &=& \gamma^2F^{\alpha}(\mathbf{r},t)F^{\beta}(\mathbf{r}',t).
\end{eqnarray}
 To distinguish the flow velocity in the equations of quantum hydrodynamics, it is necessary to substitute the explicit form of the wave function (\ref{wave3}) in the definition of the basic current density (\ref{current})
\begin{equation}\label{current2}
J^{\alpha}(\mathbf{r},t)=\rho(\mathbf{r},t) v^{\alpha}(\mathbf{r},t).
\end{equation}
The system of the evolution equations for the bosons with a short-range interaction in the absence of excited particles in terms of the particles density $\rho(\mathbf{r},t)$, the flow velocity $\mathbf{v}(\mathbf{r},t)$ and the spin density $\mathbf{F}(\mathbf{r},t)$ can be recast into a more transparent form, if we substitute the particle flux density (\ref{current2}) into the  continuity equation (\ref{nn}) and into the equation of motion (\ref{Euler}) in the self-field approximation
\begin{equation} \label{n2}
  \partial_t\rho+\partial_{\alpha}(\rho v^{\alpha})=0,
\end{equation}
\begin{widetext}
\begin{equation}\label{Euler2}
m\rho(\partial_t+v_{\beta}\partial^{\beta})v^{\alpha}+\partial_{\beta}p^{\alpha\beta}
         -\frac{\hbar^2}{4m}\partial^{\alpha}\triangle\rho
         +\frac{\hbar^2}{4m}\partial^{\beta}\biggl(\frac{\partial^{\alpha}\rho\partial^{\beta}\rho}{\rho}\biggr) =\Upsilon\rho\partial^{\alpha}\rho-\rho\partial^{\alpha}U_{ext}+\frac{\hbar^2}{2m}\rho\partial_{\beta}
         \biggl(\partial^{\alpha}\mathbf{f}\cdot\partial^{\beta}\mathbf{f} \biggr) +\gamma\mathbf{F}\partial^{\alpha}\mathbf{B} +\partial_{\beta}Q^{\alpha\beta},
\end{equation}
\end{widetext}
where the second term on the left hand side of the equation is the gradient of the kinetic pressure, which can vanish for the BEC
\begin{equation}
p^{\alpha\beta}=\biggl\langle m_ja^2u^{\alpha}_ju^{\beta}_j\biggr\rangle.
\end{equation}
The third and fourth terms on the left hand side of equation (\ref{Euler2}) comprise the quantum force density, produced by the quantum Bohm potential. The first term on the right hand side of equation (\ref{Euler2}) is the force, produced by the short-range interaction. The second term on the right hand side is the force field density due to the influence of an external potential. The third term describes the gradient of the spin part of the quantum Bohm potential or the ``spin stress''.  And the last term on the right hand side of (\ref{Euler2}) manifests the thermal-spin interactions
\begin{eqnarray}\label{Q}
Q^{\alpha\beta}&=& -\,\frac{\hbar^2}{2m}\biggl\langle a^2\partial_j^{\alpha}\mathbf{w}_j\cdot\partial_j^{\beta}\mathbf{w}_j \nonumber\\
&& +\,a^2\partial_j^{\alpha}\mathbf{w}_j\cdot\partial_j^{\beta}\mathbf{f}_j+
a^2\partial_j^{\alpha}\mathbf{f}_j\cdot\partial_j^{\beta}\mathbf{w}_j  \biggr\rangle.
\end{eqnarray}
To close the equation system (\ref{n2}), (\ref{Euler2}) we need to derive the equation for the spin-thermal force field (\ref{Q}). But, in the case of BEC we will not take into account the contribution of thermal fluctuations.

\subsection{The spin density evolution equation}

The microscopic spin density appears in the equation of motion of the bosonic fluid
\begin{equation}
  F^{\alpha}(\mathbf{r},t)=\biggl\langle \psi^{\dagger}\hat{f}^{\alpha}_j\psi \biggr\rangle
\end{equation}
and one requires an additional equation for the spin density dynamics, which can be derived by making use of the many-particle Schr\"{o}dinger equation (\ref{Shrodinger}) with the Hamiltonian (\ref{H}):
\begin{equation}
  \partial_t   F^{\alpha}(\mathbf{r},t)+\partial_{\beta}\Im^{\alpha\beta}(\mathbf{r},t)=\varepsilon^{\alpha\beta\gamma}F^{\beta}(\mathbf{r},t)B^{\gamma}(\mathbf{r},t).
\end{equation}
Here the tensor of the spin current density has the microscopic form
\begin{equation}
  \Im^{\alpha\beta}(\mathbf{r},t)=\frac{1}{2m}\biggl\langle (p^{\beta}_j\psi)^{\dagger}\hat{f}_j^{\alpha}\psi+\psi^{\dagger}\hat{f}_j^{\alpha}p_j^{\beta}\psi \biggr\rangle.
\end{equation}
In the macroscopic fluid description, the spin current density reads in terms of the fluid variables
\begin{equation} \label{spin current} \Im^{\alpha\beta}(\mathbf{r},t)=m\biggl\langle a^2f^{\alpha}_jv^{\beta}_j-\frac{\hbar}{2m}a^2\varepsilon^{\alpha\gamma\eta}f^{\gamma}_j\partial^{\beta}_jf^{\eta}_j\biggr\rangle,
\end{equation}
where the first term represents the ordinary spin current caused by the spin $\mathbf{f}_j$ transfered by a particle with the velocity $\mathbf{v}_j$, and the second term is the self-action of the spins. Finally, using the thermal decomposition and  Madelung decomposition of the $N$-particle wave function, the spin density dynamical equation is recast into
\begin{equation} \label{spin}
  \partial_t\mathbf{F}+\partial_{\beta}(\mathbf{F}v^{\beta})=
  \frac{\gamma}{\hbar}\mathbf{F}\times\mathbf{B}+\frac{\hbar}{2m}\partial_{\beta}
  \biggl( \frac{\mathbf{F}}{\rho}\times\partial^{\beta}\mathbf{F} \biggr)+\mathbf{K}.
\end{equation}
The first term of the spin density evolution equation describes the torque acting on the spin density by the magnetic field. In a self-consistent approximation of the magnetic spin-spin interactions, the magnetic field appears in the integral form
\begin{equation}
  B^{\alpha}(\mathbf{r},t)=\gamma\int d\mathbf{r}'G^{\alpha\beta}(\mathbf{r}-\mathbf{r}')F^{\beta}(\mathrm{r}',t)
\end{equation}
and satisfies the Maxwell equation
\begin{equation}   \label{Maxwell}
  \nabla\times\mathbf{B}=4\pi\gamma\nabla\times\mathbf{F}.
\end{equation}
The second term on the right hand side of the equation (\ref{spin}) represents the self-action of particle's spin, which creates the spin self-torque effect, and the last term is the thermal-spin interactions
\begin{equation}\label{L}
  \mathbf{K}=\partial_{\beta}\biggl\langle a^2z^{\beta}_j\mathbf{f}_j-a^2\mathbf{f}_j\times\partial^{\beta}_j\mathbf{w}_j \biggr\rangle.
\end{equation}
The system of equations, which contains the spin current (\ref{spin current}) and the momentum flux (\ref{Flux2}), driven by the texture of spins, was obtained for a system of a large number of interacting particles. A realistic model of a quantum spinning particle with spin-${\frac 12}$ had been derived by T. Takabayasi\cite{23,24,25}. It was shown, that even when the magnetic field is zero or the particle does not have a magnetic moment, the spin vector will experience a quantum torque, and the velocity field will be affected by the quantum torque.  The thermal corrections, characterized by the thermal-spin interactions (\ref{Q}) and (\ref{L}) appear in the equations of motion (\ref{Euler2}) and the spin density evolution (\ref{spin}). However, specifying from the general equations to the case of a Bose-Einstein condensate, we get a closed formalism of the quantum hydrodynamics, excluding thermal corrections from consideration.

\section{\label{II}Excitations in the polarized BEC}

Let us consider a system of atoms at zero temperature, which is completely in the Bose condensate state. Then we have to neglect the thermal effects (\ref{Q}) and (\ref{L}), and we are now in a position to derive the dispersion laws of the waves in the Bose condensate of cold atoms from the obtained closed system of the macroscopic equations (\ref{n2}), (\ref{Euler2}), and (\ref{spin}).

We will analyse the small perturbations in the magnetized $3D$ Bose-Einstein condensate, put in the uniform magnetic field $\mathbf{B}_0=B_0\mathbf{e}_z$. The equilibrium concentration of the cold atoms is $\rho_0$. The perturbations of equilibrium state are $\rho=\rho_0 + \delta\rho$, $\mathbf{v} = \delta\mathbf{v},$ $\mathbf{B}=\mathbf{B}_0+\delta\mathbf{B}$ and $\mathbf{F}=\mathbf{F}_0+\delta\mathbf{F}$. The small perturbations can be represented in the form $\delta f=f(\omega,k)\exp\{-i\omega t+i\mathbf{k}\mathbf{r}\}$, where the wave vector can be decomposed into two components $\mathbf{k}=k_{\perp}\mathbf{e}_{\perp}+k_{\parallel}\mathbf{e}_z$ (perpendicular and along the applied magnetic field, respectively). The tensor of magnetic permeability depends on the modulus of the wave vector
\begin{equation}
\chi^{\alpha\beta}=\begin{pmatrix} \frac{\gamma^2F_0}{\hbar}\frac{\Omega}{\omega^2-\Omega^2} & \frac{\gamma^2F_0}{\hbar}\frac{i\omega}{\omega^2-\Omega^2} & 0 \\ -\frac{\gamma^2F_0}{\hbar}\frac{i\omega}{\omega^2-\Omega^2} & \frac{\gamma^2F_0}{\hbar}\frac{\Omega}{\omega^2-\Omega^2} & 0 \\ 0 & 0 & \frac{F^2_0\gamma^2}{\rho_0m}\frac{k^2}{\omega^2-v^2_fk^2}\end{pmatrix},
\end{equation}
where the modified cyclotron frequency,
\begin{equation}
  \Omega(k)=\frac{\gamma B_0}{\hbar}+\frac{\hbar F_0}{2m\rho_0}k^2,
\end{equation}
depends on the modulus of the wave vector due to the influence of spin torque $\sim \mathbf{F}\times\partial^2\mathbf{F},$ and
\begin{equation}
  v^2_f=\frac{\hbar^2}{4m^2}k^2-\frac{\Upsilon \rho_0}{m}.
\end{equation}
Substituting these small perturbations into the system of equations (\ref{n2}), (\ref{Euler2}), (\ref{spin}) and (\ref{Maxwell}) and neglecting nonlinear terms, we obtain the dispersion equation
\begin{widetext}
\begin{equation}
\biggl(1+\frac{w_f\Omega}{\omega^2-\Omega^2}\biggr)\times\biggl(k^2+\frac{w_f\Omega k^2_{\parallel}}{\omega^2-\Omega^2}+\frac{4\pi\gamma^2 F^2_0}{m\rho_0}\frac{k^2_{\perp}}{\omega^2-v^2_fk^2}\biggr)
-\biggl(\frac{w_f\Omega}{\omega^2-\Omega^2}\biggr)^2k^2_{\parallel}=0.
\end{equation}
\end{widetext}

\subsection{Wave along to the magnetic field, $\mathbf{k}=k_{\parallel}\mathbf{e}_z$}

In the case of wave propagation parallel to an external uniform magnetic field, pure spin waves are excited with the dispersion law
\begin{equation}
  \omega=\Omega(k)=\frac{\gamma B_0}{\hbar}+\frac{\hbar F_0}{2m\rho_0}k^2
\end{equation}

\subsection{Wave perpendicular to the magnetic field, $\mathbf{k}=k_{\perp}\mathbf{e}_{\perp}$}

Consider now the waves propagating in the $xy$ plane, perpendicular to the direction of the external uniform magnetic field. In this case, several types of waves can be generated. The dispersion equation can be recast into
\begin{equation}\label{w}
  \biggl(1+\frac{\omega_f\Omega}{\omega^2-\Omega^2}\biggr)
  \biggl(1+\frac{F_0\hbar}{\rho_0 m}\frac{\omega_f}{\omega^2-v^2_fk^2}\biggr)=0,
\end{equation}
where $ \omega_f=\frac{4\pi\gamma^2F_0}{\hbar}.$  There are two solutions of the dispersion equation (\ref{w}) in view of its factorized form. The first solution yields the dispersion law
\begin{equation}\label{w1}
  \omega^2=\frac{\hbar^2}{4m^2}k^4+\frac{\rho_0}{m}\left\{|\Upsilon|-4\pi\left(\frac{\gamma F_0}{\rho_0}\right)^2\right\}.
\end{equation}
The relation (\ref{w1}) characterizes the hybrid mode, where the first term follows from the Bohm quantum potential, and the second term is characterized by the interactions for the repulsive short range interaction ($\Upsilon < 0$). As can be seen from the dispersion law (\ref{w1}), this wave can be unstable due to the influence of spin-spin interactions. But, in the most experiments, the force corrections caused by the long-range spin-spin interaction are small compared to the contribution of the short-range action. Therefore, an effective force yields a small decrease in the speed of sound, but does not violate the stability of the system.

The second solution of the dispersion equation (\ref{w}) represents the spin waves solution
\begin{equation}\label{w2}
  \omega^2=\Omega\left(\Omega-\frac{4\pi\gamma^2F_0}{\hbar}\right),\quad
  \Omega=\frac{\gamma B_0}{\hbar}+\frac{\hbar F_0}{2m\rho_0}k^2.
\end{equation}
From the formula (\ref{w2}) we can see that int long wave length limit the frequency square becomes negative, which leads to instability of spin mode.

\section{Discussion}

In this paper, we give an original derivation of the equations of quantum hydrodynamics, which are consistent with the existing models, on the basis of the collective dynamics of individual particles described by the microscopic many-particle Schrodinger equation. We therefore confirm and supplement the earlier obtained results \cite{11,26}. Starting from the many-particle Schr\"{o}dinger equation (\ref{Shrodinger}), we have constructed the quantum fluid model for the spin-1 Bose-Einstein condensate of neutral atoms, which accounts for the possible spin-spin and short-range interactions when the range of the inter-particle interactions is comparable with the radius of bosons. The essential contribution in the force field of the short-range interactions arises in the first order in the interaction radius. We demonstrate that fluid's dynamics for a spin-1 BEC is described by the set of equations, which encompasses the continuity equation (\ref{n2}), the momentum balance equation (\ref{Euler2}) and the spin density evolution equation (\ref{spin}). The force fields due to the Bohm potential and the particle spin self-interaction contribute to the momentum balance equation (\ref{Euler2}). We have shown that the spin density evolution equation is modified by a nontrivial spin torque effect. The latter arises as a result of the self-interactions between bosons' spins in the system of many-interacting atoms along with the spin-spin interactions, confirming the earlier results \cite{11,26}. But in addition, the terms responsible for the many-particle spin-spin interactions emerge in the equation of spin density evolution (\ref{spin}) and momentum balance equation (\ref{Euler2}).

In general, there are also thermal corrections in the equations of motions, characterized by the thermal-spin interactions (\ref{Q}) and (\ref{L}). However, the latter are excluded from consideration when we specialize from general equations to the case of a Bose-Einstein condensate, and we ultimately get a closed formulation of the quantum hydrodynamics. The study of thermal corrections is though an important task that requires the derivation of the equations for the force fields characterizing the spin-temperature interactions.

As an application of the formalism, we have studied the dispersion properties of the magnetized $3D$ spin-1 neutral Bose-Einstein condensate. The contribution of the equilibrium magnetization in the Bogoliubov's mode dispersion relation (\ref{w1}) is obtained. We have also derived the contribution of the spin self-interaction to the spin wave spectrum, together with the influence of an external magnetic field and the spin-spin interactions between polarized particles (\ref{w2}). It turns out that the short-range interactions do not affect the dispersion of spin waves, but the spin torque due to the self-interactions of spins leads to the dispersion of the spin waves at cyclotron frequency $\gamma B_0/\hbar$, where the square of the frequency $\sim k^2$. As we can see, the dispersion relation has a non-trivial form and in the long wave length regime the instability of the wave can arise due to the anisotropic spin-spin dipole interactions.

\subsection{Acknowledgments}
The work of Mariya Iv. Trukhanova is supported by the Russian Science Foundation  under grant 19-72-00017.

\end{document}